\documentclass[a4paper,12pt]{article}
\begin{document}
\baselineskip 0.7cm

\newcommand{\gsim}{ \mathop{}_{\textstyle \sim}^{\textstyle >} }
\newcommand{\lsim}{ \mathop{}_{\textstyle \sim}^{\textstyle <} }
\newcommand{\vev}[1]{ \left\langle {#1} \right\rangle }
\newcommand{\lsp}{ \left ( }
\newcommand{\rsp}{ \right ) }
\newcommand{\lmp}{ \left \{ }
\newcommand{\rmp}{ \right \} }
\newcommand{\llp}{ \left [ }
\newcommand{\rlp}{ \right ] }
\newcommand{\labs}{ \left | }
\newcommand{\rabs}{ \right | }
\newcommand{\EV} { {\rm eV} }
\newcommand{\KEV}{ {\rm keV} }
\newcommand{\MEV}{ {\rm MeV} }
\newcommand{\GEV}{ {\rm GeV} }
\newcommand{\TEV}{ {\rm TeV} }
\newcommand{\YR}{ {\rm yr} }
\newcommand{\mgut}{M_{GUT}}
\newcommand{\mint}{M_{I}}
\newcommand{\mgra}{M_{3/2}}
\newcommand{\mll}{m_{\tilde{l}L}^{2}}
\newcommand{\mdr}{m_{\tilde{d}R}^{2}}
\newcommand{\mllXX}[1]{m_{\tilde{l}L , {#1}}^{2}}
\newcommand{\mdrXX}[1]{m_{\tilde{d}R , {#1}}^{2}}
\newcommand{\mgy}{m_{G1}}
\newcommand{\mgl}{m_{G2}}
\newcommand{\mgc}{m_{G3}}
\newcommand{\nuR}{\nu_{R}}
\newcommand{\slL}{\tilde{l}_{L}}
\newcommand{\slLi}{\tilde{l}_{Li}}
\newcommand{\sdR}{\tilde{d}_{R}}
\newcommand{\sdRi}{\tilde{d}_{Ri}}
\newcommand{\e}{{\rm e}}
\renewcommand{\thefootnote}{\fnsymbol{footnote}}
\setcounter{footnote}{1}

\makeatletter
%
%
%
%
%
\newtoks\@stequation

\def\subequations{\refstepcounter{equation}%
  \edef\@savedequation{\the\c@equation}%
  \@stequation=\expandafter{\theequation}
  \edef\@savedtheequation{\the\@stequation}
  \edef\oldtheequation{\theequation}%
  \setcounter{equation}{0}%
  \def\theequation{\oldtheequation\alph{equation}}}

\def\endsubequations{%
  \ifnum\c@equation < 2 \@warning{Only \the\c@equation\space subequation
    used in equation \@savedequation}\fi
  \setcounter{equation}{\@savedequation}%
  \@stequation=\expandafter{\@savedtheequation}%
  \edef\theequation{\the\@stequation}%
  \global\@ignoretrue}


\def\eqnarray{\stepcounter{equation}\let\@currentlabel\theequation
\global\@eqnswtrue\m@th
\global\@eqcnt\z@\tabskip\@centering\let\\\@eqncr
$$\halign to\displaywidth\bgroup\@eqnsel\hskip\@centering
     $\displaystyle\tabskip\z@{##}$&\global\@eqcnt\@ne
      \hfil$\;{##}\;$\hfil
     &\global\@eqcnt\tw@ $\displaystyle\tabskip\z@{##}$\hfil
   \tabskip\@centering&\llap{##}\tabskip\z@\cr}

\makeatother


\begin{titlepage}

\begin{flushright}
KEK-TH-774

UT-949
\end{flushright}

\vskip 0.35cm
\begin{center}
{\large \bf  Ultra High Energy Cosmic Ray, Superheavy Dark Matter and
 Extra Dimension}
\vskip 1.2cm
Kaoru Hagiwara$^{a)}$ and Yosuke Uehara$^{b)}$

\vskip 0.4cm

$^{a)}$ {\it Theory Group, KEK, Tsukuba, Ibaraki 305-0801, Japan}

$^{b)}$ {\it Department of Physics, University of Tokyo, 
         Tokyo 113-0033, Japan}\\
\vskip 1.5cm

\abstract{We propose a new mechanism for explaining the very long lifetime of
superheavy dark matter $X$, which is proposed as a
source of the ultra high-energy cosmic rays above the
GZK cutoff ($5 \times 10^{19} \EV $). The singlet $X$ particle
couples to the MSSM particles only through a bulk singlet field
which develops the v.e.v. in the ``hidden'' brane.
The distance between this hidden brane and the ``visible'' brane
naturally leads to the exponential suppression of the coupling.
The $X$ particle decays predominantly into the higgsino and higgs boson
of the MSSM, and its decay spectrum is completely determined once their
properties are known.}

\end{center}
\end{titlepage}

\renewcommand{\thefootnote}{\arabic{footnote}}
\setcounter{footnote}{0}

%
%
%
%

\section{Introduction}

Some experiments have observed \cite{AGASA,FLYSEYE,HAVERAH,SUGAR} 
cosmic rays whose energy are above the GZK cutoff 
($5 \times 10^{19} \EV$) \cite{GZK}. 
The existence of such ultra high energy cosmic rays(UHECR) is a great puzzle 
not only for astrophysics but also for particle physics \cite{BS}.

Many scenarios have been proposed to solve this puzzle. 
``Top-Down'' scenarios typically assumes some superheavy objects
like topological defects with very long lifetime \cite{BR,H,BKV},
 whose decay products are responsible for the observed UHECR.

It has also been proposed that UHECR may be produced 
from the decay of superheavy darkmatter $X$ \cite{KR, FKN, BKV}.
Such heavy quasi-stable particles may not overclose the universe \cite{GK}
if they are gravitationally generated by inflation during the reheating
epoch just after the end of inflation \cite{CKR,KT}.

As an origin of UHECR, its mass $m_{X}$, 
its lifetime $\tau_{X}^{}$ and its abundance $\Omega_{X} h^{2}$
must satisfy specific conditions. It must be heavy enough to explain
the energy of UHECR, it must survived until now, and
the flux of $X$ decay must be consistent with observation.
These conditions lead to the constraints \cite{BS} 
\begin{subequations}
\begin{eqnarray}
& m_{X}^{} & \gsim  10^{12} \GEV, \\
10^{10} \YR \lsim & \tau_{X}^{} & \lsim 10^{22} \YR, \label{lifetime} \\
10^{-12} \lsim & \Omega_{X} h^{2} & \lsim 1.
\end{eqnarray}
\end{subequations}
Hence in order to realize this scenario, we should find a mechanism 
to make the lifetime of $X$ long enough;
\begin{eqnarray}
10^{56} \lsim m_{X}^{} \tau_{X}~{} \lsim 10^{66}.
\end{eqnarray}

Many mechanisms to realize this long but finite lifetime are proposed.
One may restrict the decay of $X$ by a discrete gauge symmetry \cite{HNY}, 
or one may assume that $X$ is perturbatively stable but that it decays via 
non-perturbative instanton effects \cite{KR} or by 
quantum gravity effects \cite{BKV}.

In this paper, we propose a new mechanism to stabilize
the superheavy object,where its decay width is exponentially suppressed 
because of the separation between the visible brane and the hidden brane.

\section{The Model}

We assume that our world is higher-dimensional(4+n dimension), and 
that it has at least two 3-branes. 
We are confined on one 3-brane (``visible brane''). 
In addition to our brane, another 3-brane (``hidden brane'') exists.
Both the MSSM particles and the superheavy dark matter $X$ are confined
in the visible brane, whereas a gauge-singlet field $\phi$ propagates
in the bulk.

We impose a $Z_{2}$ symmetry to forbid direct couplings between $X$ and
the MSSM particles in our brane. The $Z_{2}$ charge 
of the MSSM particles is 0, and that of $X$ and $\phi$ is 1. 
Then the lowest-dimensional superpotential which is
relevant for the decay of $X$ is written as
\begin{eqnarray}
W=\frac{1}{M_{*}} \phi X H_{u} H_{d}.
\end{eqnarray}
Where $H_{u}$ and $H_{d}$ stands for the higgs superfield of the MSSM.
Let us now suppose that $\phi$ has a vacuum expectation value on
the hidden brane. Naively, its v.e.v. should be of the order of the 
fundamental scale $M_{*}$.
But since it appears in the hidden brane, the v.e.v. which
we feel on the visible brane is not $M_{*}$. Let $x$ be the four-dimensional
coordinates, and $y$ be the extra-dimensional coordinates. $y=0$ 
denotes the location of the visible brane, and $y=y_{*}$ denotes that
of the hidden brane. The v.e.v. $\vev{\phi}$ depends only on the y
coordinates, and it satisfies
\begin{eqnarray}
\vev{\phi} (y) = \vev{\phi} (y_{*}) \times \Delta_{n} (|y-y_{*}|),
\end{eqnarray}
where
\begin{eqnarray}
\Delta_{n}(r) = (-(\partial^2)_{n} + m_{\phi}^{2})^{-1}(r) \propto \int d^{n} k e^{ik \cdot y} \frac{1}{k^2+m_{\phi}^{2}},
\end{eqnarray}
if the extra dimensional space is large enough as compared to the
inter-brane distance $r$.

The suppression factor depends on the number of extra dimensions.
It has the following simple form at long distances ($m_{\phi} r \gg 1$) 
\cite{AD,ADDM}:
\begin{subequations}
\begin{eqnarray}
\Delta_{1}(r) &=& e^{- m_{\phi} r}, \\
\Delta_{2}(r) & \sim & \frac{e^{-m_{\phi} r}}{\sqrt{m_{\phi} r}},  \\
\Delta_{n}(r) & \sim & \frac{e^{-m_{\phi} r}}{(m_{\phi} r)^{n-2}} \ (n \ge 3).
\end{eqnarray}
\end{subequations}
Thus, if the distance $r$ between our brane and hidden brane
is sufficiently large, the superpotential relevant for the decay of $X$ becomes
\begin{eqnarray}
W = \frac{\vev{\phi}(y_{*})}{M_{*}} \Delta_{n}(r) X H_{u} H_{d} \equiv g_{{\rm eff}} X H_{u} H_{d}. \label{superpotential}
\end{eqnarray}
The effective coupling constant $g_{{\rm eff}}$ is now 
suppressed exponentially, and hence it can be extremely
small without any fine tuning. The lifetime of the X-boson($\tau_{X}$)
and that of the X-fermion($\tau_{\tilde{X}}$) are essentially determined 
by the superpotential (\ref{superpotential}) because the soft SUSY breaking effects
for the effective X couplings are suppressed by powers of
$\frac{m_{SUSY}}{M_{*}}$.
In the so-called decoupling limit where the soft-SUSY breaking mass
terms are larger than the electroweak symmetry breaking scale, we find
\begin{eqnarray}
\tau_{X}^{-1}=\tau_{\tilde{X}}^{-1}=\frac{(g_{{\rm eff}} \cos \theta_{{\rm X}}^{})^{2}}{2 \pi} m_{X},
\end{eqnarray}
where $\cos \theta_{{\rm X}}^{}$ is a mixing angle of the singlet sector
and we denote the lighter mass eigenstate as X for brevity.
The long lifetime (\ref{lifetime}) required for the UHECR
candidate is satisfied when
\begin{eqnarray}
1.15 \times 10^{-33} (\frac{10^{13} \GEV}{m_{X}})^{1/2} \lsim g_{{\rm eff}} \cos \theta_{{\rm X}}^{} \lsim 1.15 \times 10^{-27} (\frac{10^{13} \GEV}{m_{X}})^{1/2},
\end{eqnarray}
For $m_{X} \sim 10^{13} \GEV$ ,$\vev{\phi} (y_{*}) \sim M_{*}$ and
 $\cos \theta_{{\rm X}}^{} \sim \frac{1}{\sqrt{2}}$, 
the required suppression is achieved by
\begin{subequations}
\begin{eqnarray}
62 \lsim & m_{\phi} r & \lsim 75 \ \ \ \ (n=1), \\
60 \lsim & m_{\phi} r & \lsim 73 \ \ \ \ (n=2), \\
62-4(n-2) \lsim & m_{\phi} r & \lsim 75-4.2(n-2) \ \ \ \ (n \ge 3).
\end{eqnarray}
\end{subequations}

\section{Experimental Signals}

The remarkable feature of this scenario is that the superheavy dark matter
$X$ decays mainly into $H_{1}$ and $H_{2}$, the MSSM higgs bosons and
their superpartners. Therefore once the higgs sector of
the MSSM is experimentally determined (once the masses and the mixing among
the higgs particles and the gauge particles are known), 
then the X decay spectrum is completely determined.

In the decoupling limit, the decay patterns are especially simple.
If X is a boson, its decay branching ratios are
\begin{eqnarray}
B(\tilde{\phi}_{u}^{+} \tilde{\phi}_{d}^{-}) = B(\tilde{\phi}_{u}^{-} \tilde{\phi}_{d}^{+}) = \frac{1}{2} B(\tilde{\phi}_{u}^{0} \tilde{\phi}_{d}^{0}) = \frac{1}{4}.
\end{eqnarray}
Whereas if X is a fermion, they are
\begin{subequations}
\begin{eqnarray}
B(W^{+} \tilde{\phi}_{d}^{-}) = B(W^{-} \tilde{\phi}_{d}^{+}) = B(H^{+} \tilde{\phi}_{u}^{-}) = B(H^{-} \tilde{\phi}_{u}^{+}) &=& \frac{1}{8} \sin^{2} \beta, \\
B(W^{+} \tilde{\phi}_{u}^{+}) = B(W^{-} \tilde{\phi}_{u}^{+}) = B(H^{+} \tilde{\phi}_{d}^{-}) = B(H^{-} \tilde{\phi}_{d}^{+}) &=& \frac{1}{8} \cos^{2} \beta, \\
B(Z \tilde{\phi}_{d}^{0}) = B(h \tilde{\phi}_{d}^{0}) = B(H \tilde{\phi}_{u}^{0}) = B(A \tilde{\phi}_{u}^{0}) &=& \frac{1}{8} \sin^{2} \beta, \\
B(Z \tilde{\phi}_{u}^{0}) = B(h \tilde{\phi}_{u}^{0}) = B(H \tilde{\phi}_{d}^{0}) = B(A \tilde{\phi}_{d}^{0}) &=& \frac{1}{8} \cos^{2} \beta. 
\end{eqnarray}
\end{subequations}
Here $\tilde{\phi}$ denotes the gauge eigenstates of the higgsinos,
and $h,H,A,H^{\pm}$ are the MSSM higgs bosons.
In the decoupling limit, the lighter CP-even higgs boson h
reduces to the SM higgs boson, and all the remaining
higgs bosons and higgsinos are degenerate.
The W and Z bosons are longitudinally polarized.

In the real world, the mass eigenstates are the mixtures of the
higgs bosons, the higgsinos and the gauginos of the same spin and charge.
Nevertheless, because the X mass is far greater than any of the 
MSSM particles, the above decay branching fractions remain valid
simply by replacing the current eigenstate as appropriate summation over
the mass eigenstate contributions.
W and Z decay properties are known, and we expect $h,H,A,H^{\pm}$ 
decays to contain b and t quarks, The observed UHECR signal may be
explained by protons and neutrons from these quark jets. The decay
patterns of charginos and neutralinos (mass eigenstates of charged and
neutral colorless SUSY fermion) depend more strongly on details of the
SUSY breaking mechanism. It is likely, however, that their decays also
contain W and Z bosons as well as b and t quarks.
We should expect significant amount of neutrinos accompanying the UHECR
events. If R-parity is conserved, then the bulk of the decaying X energy 
may be carried by the lightest supersymmetric particle(LSP).
 
\section{Summary}

In this paper, we propose a new mechanism to stabilize the superheavy dark
matter which may be the origin of the observed ultra high energy cosmic ray.

In our scenario, the long lifetime of the superheavy darkmatter $X$ is
realized by the separation between the visible brane and the hidden
brane in a large extra dimensional space.

$X$ decays mainly into higgs and higgsino, so
this scenario may be testable from the energy spectrum of decayed
products. In the future it may be possible to directly detect
ultra high energy neutrinos or neutralinos(LSP). 

\noindent
{\bf Acknowledgment}

We thank K.Hamaguchi and T.Yanagida for stimulating discussions.

\noindent
{\bf note added}

After we finished this paper, we learned from S.Sarkar that 
he and his collaborators proposed\cite{EGLNS,GGS} that cryptons 
- bound states of the fractional charges which arise in the 
massless spectrum of the heterotic string compactifications 
- may constitute the dark matter, and that they may account 
for the UHECR.  In this scenario, the superheavy crypton lives 
in the hidden sector and it decays only through higher-order 
non-resnomalizable operators.  We also learned that cosmic ray 
spectrum from the decays of superheavy objects has been 
studied in detail\cite{BS,S} by using the HERWIG event 
generator\cite{HERWIG}.

%
%
\newcommand{\Journal}[4]{{\sl #1} {\bf #2} {(#3)} {#4}}
\newcommand{\PL}{\sl Phys. Lett.}
\newcommand{\PR}{\sl Phys. Rev.}
\newcommand{\PRL}{\sl Phys. Rev. Lett.}
\newcommand{\NP}{\sl Nucl. Phys.}
\newcommand{\ZP}{\sl Z. Phys.}
\newcommand{\PTP}{\sl Prog. Theor. Phys.}
\newcommand{\NC}{\sl Nuovo Cimento}
\newcommand{\PAN}{\sl Phys.Atom.Nucl.}

\end{document}